\documentclass[5p, twocolumn]{elsarticle}

\usepackage{dcolumn}
\usepackage{bm, amssymb, color}
\usepackage{hyperref, lineno}
\usepackage{natbib}
\usepackage{tabulary}
\usepackage{multirow}
\usepackage{amsmath,soul}
\usepackage{amsfonts}
\usepackage{bbm}
\usepackage[export]{adjustbox}
\usepackage{float}
\usepackage[caption=false]{subfig}
\usepackage{array}
\usepackage{cases}
\usepackage{makecell}
\usepackage{physics}
\usepackage{yhmath}
\modulolinenumbers[5]
\journal{Acta Materialia}

\usepackage[colorinlistoftodos,shadow,textwidth=18mm]{todonotes}

\newcommand{\fn}[2]{\mathinner{#1\mathopen{\left(#2\right)}}}
\newcommand{\spD}[1]{\fn{\tilde{\chi}_{_V}}{#1}}
\newcommand{\vect}[1]{\bm{#1}}

\newcommand{\K}{\mathbbm{k}}

\usepackage{soul,xcolor}
\setstcolor{red}
\newcommand{\E}[1]{\left\langle#1\right\rangle}
\usepackage[colorinlistoftodos,shadow,textwidth=18mm]{todonotes}

\begin{document}

\begin{frontmatter}

\title{Microstructural and Transport Characteristics of Triply Periodic  Bicontinuous Materials }

\author[chem,phys,prism,appmath]{Salvatore Torquato\corref{mycorrespondingauthor}}
\cortext[mycorrespondingauthor]{Corresponding author}
\ead{torquato@princeton.edu}
\ead[url]{https://torquato.princeton.edu/}

\author[chem,prism]{Jaeuk Kim}
\address[chem]{Department of Chemistry, Princeton University, Princeton, New Jersey 08544, USA}
\address[phys]{Department of Physics, Princeton University, Princeton, New Jersey 08544, USA}
\address[prism]{Princeton Materials Institute, Princeton University, Princeton, New Jersey 08544, USA}
\address[appmath]{Program in Applied and Computational Mathematics, Princeton University, Princeton, New Jersey 08544, USA}

\newtheorem{conj}{Conjecture}

\begin{abstract}

  Three-dimensional (3D) bicontinuous two-phase materials are increasingly gaining interest because of their unique
multifunctional characteristics and advancements in techniques to fabricate them.
Because of their complex topological and structural properties, it still has been nontrivial
to develop explicit microstructure-dependent formulas to predict accurately their physical properties. A primary goal of the present paper is to ascertain various microstructural and transport characteristics
of five different models of triply periodic bicontinuous porous materials at a porosity $\phi_1=1/2$:
those in which the two-phase interfaces are the Schwarz P, Schwarz D and Schoen G minimal surfaces as well as two different pore-channel structures. 
We ascertain their spectral densities, pore-size distribution functions, local volume-fraction variances, and  hyperuniformity order metrics and then use this information to estimate certain effective steady-state as well as time-dependent transport properties via closed-form microstructure-property formulas. 
Specifically, the recently introduced time-dependent diffusion spreadability is determined exactly from the spectral density. 
Moreover, we accurately estimate the fluid permeability of such porous materials
from a closed-form formula that depends on the second moment
of the pore-size function and the formation factor, a measure of the tortuosity
of the pore space, which is exactly obtained for the three minimal-surface structures. 
We also rigorously bound the permeability from above using the spectral density.  For the five models with identical cubic unit cells,  we find that the permeability, inverse of the specific surface, hyperuniformity order metric, pore-size second moment
and long-time spreadability behavior are all positively correlated and rank order the structures in exactly the same way. 
We also conjecture what structures maximize the fluid permeability for arbitrary porosities and show that this conjecture must be true in the extreme porosity limits by identifying the corresponding optimal structures.

\end{abstract}

\end{frontmatter}

\section{Introduction}

Two-phase heterogeneous materials (media) abound in Nature and synthetic situations. 
Examples of such materials include composites and porous media, biological media, foams, polymer blends, granular media, cellular solids, geological media, and colloids \citep{To02a,Mi02,Sa03,Zo05}.
It is well-established that the effective properties of composites generally depend on an infinite set of correlation functions that fully characterize the microstructure \cite{To02a}. 
Since such complete information is generally not available, it is useful to devise estimates of the effective properties that depend on nontrivial microstructural information, including accurate approximation formulas \cite{To02a} and rigorous bounds \cite{To02a,Mi02}.
Concerning the latter approach, it is well known that the  microstructures that maximize or minimize the effective electrical (thermal) conductivities as well as bulk moduli of macroscopically isotropic two-phase composites
at a fixed volume fraction consist of a topologically {\it disconnected} phase dispersed throughout a {\it continuous} (percolating) matrix phase \cite{To02a,Mi02}.
These extremal structures include the Hashin-Shtrikman sphere assemblages \cite{Ha62c}, certain laminates \cite{Lu85,Mi86b} and
Vigdergauz constructions \cite{Vig89}.

\begin{figure*}[bthp]
\centering
\includegraphics[width=1\textwidth]{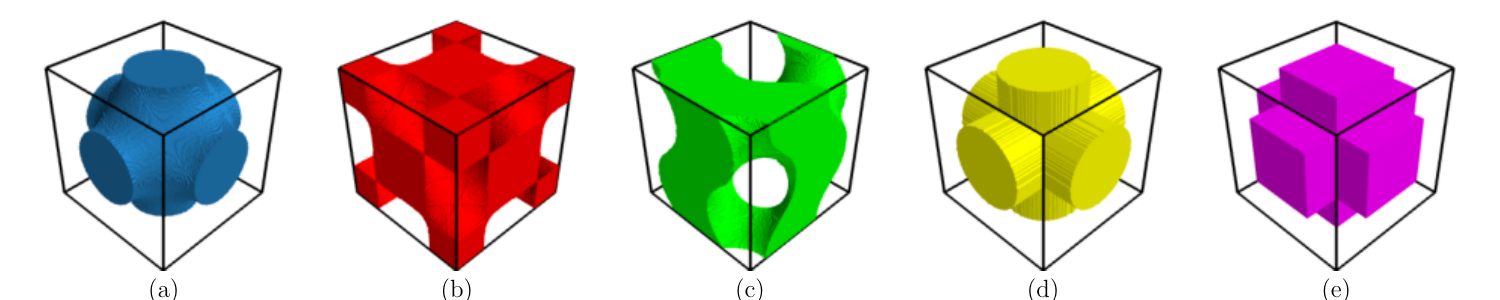}
\caption{
Fundamental (unit) cells of phase 1 domains for the five triply periodic models with porosity $\phi_1=0.5$ that are considered in
this work: (a) Schwarz P, (b) Schwarz D, (c) Schoen G, (d) spherical-pore/circular-channel model, and (e) cubic-pore/square-channel model.
For each of the three triply periodic minimal surfaces (a-c), the domains of both percolating phases are identical up to simple
translation and reflection transformations, implying that they obey statistical phase-inversion symmetry; see \ref{app:phase-inversion}. 
The general spherical-pore/circular-channel model has a single spherical pore of radius $a+b$ and three perpendicular cylindrical channels of radius $a$, which intersect at the center of the fundamental cell; see Ref.
\cite{Ju05, Ge09a}.
Analogously, the general cubic-pore/square-channel model consists of a cubic
pore of side length $2(a+b)$ and three square channels of side length
$2a$.
We study these two pore-channel models with $b=0$, called the circular- and square-channel models.
\label{fig:models}
}
\end{figure*}

It is common for two-phase media to be bicontinuous in three-dimensional Euclidean space $\mathbb{R}^3$.
 A {\it bicontinuous} composite is one in which both phases of a two-phase composite are topologically connected across the sample.
This topological feature, i.e., percolation of both phases, is
rare in two dimensions, while very common in three dimensions \cite{To02a}.
Bicontinuous media that are periodic in three dimensions, i.e., {\it triply periodic},\footnote{More precisely, triply periodic
media possess fundamental cells  that periodically fill all of the three-dimensional Euclidean space $\mathbb{R}^3$ and possess the symmetry of
one of the crystallographic space groups.} are an important class of two-phase media. 
They are increasingly gaining interest because of their desirable physical properties and a capacity to readily fabricate them due to advancements in additive manufacturing \cite{Va13}.
Within this class, it has been shown that certain triply periodic minimal 
surfaces are optimal for several types of multifunctional performance \cite{To02d,To03b,To04b,Ju05,Ge09a}.
Triply periodic minimal surfaces  in $\mathbb{R}^3$ \cite{An90,Mac93,Kl96,Ju97,Lo03}, which arise in 
a multitude of physical \cite{Ol98,Lu01,Zi00,Ka11} and biological contexts \cite{Gel94,La95,Na96,Ka11}, are those in which the mean curvature is everywhere zero.\footnote{
The mean curvature $H({p})$ at a point ${p}$ on a surface in three-dimensional space
is the average of the two principal curvatures $\kappa_1({p})$ and $\kappa_2({p})$, i.e.,
$H({p})= [\kappa_1({p})+\kappa_2({p})]/2$, vanishes at every point $p$ on the surface, implying that the principal curvatures have the same magnitude but opposite signs and hence each ${p}$ is a saddle point.}
Examples of such surfaces include the Schwarz primitive (P), the Schwarz diamond (D), and the Schoen gyroid (G) surfaces within their fundamental periodic cells are shown in  Fig. \ref{fig:models}, among other triply periodic bicontinuous media that we consider in this paper, namely, ``pore-channel" models.
The triply periodic minimal surfaces shown in Fig. \ref{fig:models} partition space into two disjoint but intertwining regions that are simultaneously continuous in which the phase volume fractions $\phi_1$ and $\phi_2$ are identical, i.e., $\phi_1=\phi_2=1/2$.
In \ref{app:phase-inversion}, we show that the domains of both percolating phases in the P, D, and G minimal surfaces are identical up to simple translation and reflection transformations, implying that they obey  {\it statistical phase-inversion symmetry}.

It has been demonstrated that triply periodic two-phase bicontinuous composites with interfaces that are the Schwarz  P and D minimal
surfaces are not only geometrically extremal but extremal 
when heat transport competes with electrical transport of heat and electricity \cite{To02d,To03b}. 
More specifically, these triply periodic minimal surfaces maximize the values of the sum of the effective thermal conductivity $\lambda_e$ and electrical conductivity $\sigma_e$ of three-dimensional two-phase composites at 50\% volume fraction with symmetric ``ill-ordered" phases, i.e., when the heat conductivity phase contrast ratio is the inverse of the electrical conductivity phase contrast ratio.
Moreover, such triply periodic composites have also been discovered to be optimal for certain multifunctional bulk modulus and electrical conductivity optimizations \cite{To04b}.

Furthermore, the macroscopically isotropic porous medium with the Schwarz P interface, where phase 1 is the pore phase and hence has a porosity $\phi_1=1/2$,  was found to have the largest fluid
permeability $\K$ \cite{Ju05} as well as the largest mean survival time $\tau$ \cite{Ge09a}, among a wide class of  
triply periodic porous media that were examined in these studies. The former study led to the following conjecture:
\begin{conj} \label{conj1}
    Among three-dimensional porous media at porosity $\phi_1=1/2$ within a simple cubic fundamental cell of side length $L$ under periodic boundary conditions, the dimensionless isotropic fluid permeability $\K/L^2$ is maximized for the structure that minimizes the total interface area, which is proposed to be the  Schwarz primitive (P) minimal surface \cite{Ju05}.
\end{conj}
\smallskip
Subsequently, strong numerical evidence (using level-set methods)  was provided to support
the proposition that the Schwarz P surface is the structure that  minimizes the total interface area \cite{Ju97}.

Due to the complexity of the topological and structural
properties of  triply periodic bicontinuous materials, it has been nontrivial to 
develop explicit microstructure-dependent formulas
that enable accurate predictions of their physical properties.
Given the importance of this class of materials, a primary goal of the present paper is to ascertain various microstructural characteristics, including the spectral densities, pore-size distribution functions, local volume-fraction fluctuations, and the associated hyperuniformity order metrics for the structures shown in Fig. \ref{fig:models}. 
Some of this microstructural information is then used to estimate certain effective steady-state as well as time-dependent transport properties via closed-form microstructure-property formulas. 
The steady-state properties examined are the macroscopically isotropic fluid permeability $\K$ and the effective electrical conductivity $\sigma_e$.\footnote{Due to the cubic symmetry, both the fluid permeability and effective conductivity are scalar quantities, i.e., the corresponding tensors are proportional to the identity tensor, and thus the properties are macroscopically isotropic.}  
We also determine the recently introduced time-dependent diffusion spreadability ${\cal S}(t)$ as a function of time $t$. 
Importantly, among these physical properties, the most challenging to predict theoretically for general microstructures is the fluid permeability $\K$, which is defined by Darcy's law \cite{Sc74,To02a,Sa03}. 
We also generalize {\bf Conjecture 1} for $\phi_1=1/2$ to include the structures that maximize the fluid permeability for arbitrary porosities, and then show that this generalized conjecture must be true in the extreme porosity limits ($\phi_1$ tending to zero and to unity) by identifying the corresponding optimal structures.

In Sec. \ref{defs}, we present pertinent background/definitions of the microstructural descriptors and hyperuniformity concept.
In Sec. \ref{transport}, explicit formulas that relate transport properties to the microstructure are provided and discussed, including the derivation of a formula for the effective conductivity $\sigma_e$ that applies to general bicontinuous media, which reduces to an exact for $\sigma_e$ for the triply periodic minimal surfaces and is an excellent approximation for $\sigma_e$ of general bicontinuous media for $\phi_1$ in the vicinity of $1/2$. 
We also provide relevant known formulas for the fluid permeability and the spreadability 
that are functionals of certain statistical descriptors, which are computed in Sec. \ref{micro}.
In Sec. \ref{estimates}, we apply the microstructure-dependent formulas of Sec. \ref{transport} and results of  Sec. \ref{micro}  to predict the aforementioned transport properties of the five triply periodic bicontinuous models shown in Fig. \ref{fig:models}. 
In Sec. \ref{max-perm}, we generalize {\bf Conjecture \ref{conj1}} to include the structures that maximize the permeability for all nontrivial porosities, i.e., $0 < \phi_1 <1$. Finally, in Sec. \ref{discuss}, we remark on how the various structural and physical properties of the five triply periodic media models studied here are positively correlated with one another. We also describe open problems.

\section{Background on Microstructural Descriptors and Hyperuniformity}
\label{defs}

There are a multitude of different statistical descriptors of the microstructure
of two-phase media; see Ref. \citep{To02a} and references therein.
The most relevant for the purposes of this paper are the
$n$-point correlation functions, spectral density, pore-size distributions, local volume-fraction fluctuations,
and hyperuniformity characteristics, which are defined below and applied in subsequent sections.

\subsection{$n$-Point Correlation Functions}
\label{n-point}

A two-phase random medium is a domain of space $\mathcal V \subseteq \mathbb{R}^d$ that is partitioned into two disjoint regions that make up $\mathcal V$: a phase 1 region $\mathcal V_1$ of volume fraction $\phi_1$ and a phase 2 region $\mathcal V_2$ of volume fraction $\phi_2$ \citep{To02a}. The phase indicator function $\mathcal I^{(i)}(\mathbf x;\omega)$ for a given realization $\omega$ is defined as
\begin{equation} \label{indicator}
\mathcal I^{(i)}(\mathbf x;\omega) \equiv
\begin{cases}
    1, & \mathbf x \in \mathcal V_i,\\ 
    0, & \mathbf x \notin \mathcal V_i,
\end{cases}
\qquad (i=1,2).
\end{equation}
The statistical properties of each phase of the system are specified by the countably infinite set of $n$-point correlation functions $S_n^{(i)}$, which are defined by \citep{To82b, St95,To02a}:
\begin{eqnarray}\label{Sndef}
S_n^{(i)}(\mathbf{x}_1, \ldots, \mathbf{x}_n) \equiv  \left\langle\prod_{j=1}^n {\cal I}^{(i)}(\mathbf{x}_j;\omega)\right\rangle,
\end{eqnarray}
where angular brackets denote an ensemble average over realizations. The $n$-point correlation function $S_n^{(i)}(\mathbf{x}_1, \ldots, \mathbf{x}_n)$ has a probabilistic interpretation: it gives the probability of finding the ends of the vectors $\mathbf{x}_1, \ldots, \mathbf{x}_n$ all in phase $i$.
For this reason,  $S_n^{(i)}(\mathbf{x}_1, \ldots, \mathbf{x}_n)$ is sometimes referred to as the  {\it $n$-point probability function}.

If the medium is statistically homogeneous, $S_n^{(i)}(\mathbf{x}_1, \ldots, \mathbf{x}_n)$ is translationally invariant
and, in particular, the one-point correlation function is independent of position and equal to  the volume fraction of phase $i$:
\begin{eqnarray}
S_1^{(i)}(\mathbf{x}) = \phi_i,
\label{S1}
\end{eqnarray}
while the two-point correlation function $S_2^{(i)}({\bf r})$ depends on the displacement vector
${\bf r} \equiv \mathbf{x}_2 -\mathbf{x}_1$. 

For statistically homogeneous media, the \textit{autocovariance} function $\fn{\chi_{_V}}{\vect{r}}$ can be defined in terms of the mean-zero fluctuating indicator function, 
\begin{equation}
\fn{\mathcal{J}^{(i)}}{\vect{r}}\equiv \fn{\mathcal{I}^{(i)}}{\vect{r}} - \phi_i, 
\label{J}
\end{equation}
as follows \cite{To02a}:
\begin{equation}
\fn{\chi_{_V}}{\vect{r}} \equiv \langle {\fn{\mathcal{J}^{(i)}}{\vect{r}'}\fn{\mathcal{J}^{(i)}}{\vect{r}'+\vect{r}} }\rangle,
\end{equation}
which is identical for each phase.
At the extreme limits of its argument, $\fn{\chi_{_V}}{\vect{r}}$ has the following asymptotic behaviors:
\begin{equation}
\chi_{_V}({\bf r}=0)=\phi_1\phi_2, \qquad \lim_{|{\bf r}| \rightarrow \infty} \chi_{_V}({\bf r})=0,
\label{limits}
\end{equation}
in which the latter limit applies when the medium possesses no long-range order. If the medium is statistically homogeneous and isotropic, then 
${\chi}_{_V}({\bf r})$ depends only on the magnitude of its argument $r=|\bf r|$,
and hence is a radial function. In such instances, its slope at the origin is directly related
to the {\it specific surface} $s$ (interface area per unit volume); specifically, we have in any space
dimension $d$, the small-$r$ asymptotic form \cite{To02a},
\begin{equation}
\chi_{_V}(r)= \phi_1\phi_2 - \beta(d) s r + {\cal O}(r^2),
\label{specific}
\end{equation}
where
\begin{equation}
\beta(d)= \frac{\Gamma(d/2)}{2\sqrt{\pi} \Gamma((d+1)/2)}
\label{beta}
\end{equation}
and $\Gamma(x)$ is the gamma function.

\subsection{Spectral Density}
\label{sec:spectral_density}

A microstructural quantity of key interest in this paper is the spectral density ${\tilde \chi}_{_V}({\bf k})$, 
which  is  the Fourier transform of $\chi_{_V}({\bf r})$, i.e.,
\begin{equation}
{\tilde \chi}_{_V}({\bf k}) = \int_{\mathbb{R}^d} \chi_{_V}({\bf r}) \,e^{-i{\bf k \cdot r}} {\rm d} {\bf r} \ge 0, \qquad \mbox{for all} \; {\bf k}.
\label{def-spec}
\end{equation}
It is a nonnegative function for all wavevectors $\bf k$ and can be obtained experimentally from scattering intensity measurements \citep{De49,De57}.
For a general statistically homogeneous two-phase medium, the spectral density must obey the following sum rule \cite{To02a,To20}:
\begin{equation}
\frac{1}{(2\pi)^d}\int_{\mathbb{R}^d} {\tilde \chi}_{_V}({\bf k})\, d{\bf k}= \chi_{_V}({\bf r}=0)=\phi_1\phi_2.
\label{sum}
\end{equation}
This follows immediately from the Fourier representation of the autocovariance function, i.e.,
\begin{equation}
\chi_{_V}({\bf r}) = \frac{1}{(2\pi)^d} \int_{\mathbb{R}^d} {\tilde \chi}_{_V}({\bf k})\, e^{i{\bf k \cdot r}} {\rm d} {\bf k}.
\end{equation}
For statistically isotropic media, the spectral density only depends on the wavenumber $k=|{\bf k}|$ and, as a consequence of  Eq. \eqref{specific}, its decay in the large-$k$ limit is controlled by the following exact power-law form:
\begin{equation}
{\tilde \chi}_{_V}({\bf k}) \sim \frac{\gamma(d)\,s}{k^{d+1}}, \qquad k \rightarrow \infty,
\label{decay}
\end{equation}
where $s$ is the specific surface and
\begin{equation}
\gamma(d)=2^d\,\pi^{(d-1)/2} \,\Gamma((d+1)/2) \,\beta(d)
\end{equation}
is a $d$-dimensional  constant and $\beta(d)$ is given by (\ref{beta}).

\subsection{Volume-Fraction Fluctuations and Hyperuniformity}

A hyperuniform point configuration in $d$-dimensional Euclidean space
$\mathbb{R}^d$ is one in which there is an anomalous suppression
of large-scale density fluctuations compared to ordinary disordered  systems,
such as typical liquids \cite{To03a,To18a}, as defined by
a structure factor $S({\bf k})$ that vanishes as the wavenumber $k\equiv |\bf k|$ tends to zero, i.e.,
\begin{equation}
\lim_{|{\bf k}| \to 0} S({\bf k})=0.
\end{equation}
All perfect crystals and many perfect quasicrystals are hyperuniform. Moreover,
there are special disordered systems that are hyperuniform.  They are exotic ideal states of amorphous matter that have attracted
great attention because they have characteristics that lie
between a crystal and a liquid;  they are like perfect crystals in the way they suppress large-scale density fluctuations
and yet are like liquids or glasses in that they are statistically isotropic with no Bragg peaks
and hence lack any conventional long-range order \citep{To18a}. These unusual attributes can endow disordered hyperuniform systems 
with novel optical, mechanical, and transport properties \citep{To18a,Ding18}. 

The hyperuniformity concept was generalized to the case of two-phase heterogeneous materials \citep{Za09}
by considering the large-$R$ behavior of local variance $\sigma_{_V}^2(R)$ associated with volume-fraction
fluctuations within a spherical window of radius $R$. Generally, $\sigma_{_V}^2(R)$ 
is related to the autocovariance function as follows \citep{Lu90b}:
\begin{eqnarray}
\sigma_{_V}^2(R) = \frac{1}{v_1(R)} \int_{\mathbb{R}^d} \chi_{_V}(\mathbf{r}) \alpha_2(r; R) d\mathbf{r},
\label{phi-var-1}
\end{eqnarray}
where $v_1(R)$ is the volume of a sphere of radius $R$, $\alpha_2(r;R)$ is the intersection volume of two identical spheres of radius $R$ (scaled by the volume of a sphere) whose centers
are separated by a distance $r$, which is known analytically in any space dimension \citep{To03a,To06b}.
Alternatively,  there is a Fourier representation of $\sigma_{_V}^2(R)$ in terms of the spectral density ${\tilde \chi}_{_V}(\mathbf{k})$ \cite{Za09}:
\begin{eqnarray}
\sigma_{_V}^2(R) = \frac{1}{v_1(R)(2\pi)^d} \int_{\mathbb{R}^d} {\tilde \chi}_{_V}(\mathbf{k}) {\tilde \alpha}_2(k; R) \dd{\mathbf{k}},
\label{phi-var-2}
\end{eqnarray}
where 
\begin{equation}
{\tilde \alpha}_2(k;R)=(4\pi)^{d/2} \Gamma(1+d/2)\frac{J^2_{d/2}(kR)}{k^d}
\label{alpha-tilde}
\end{equation}
 is the Fourier transform of $\alpha_2(r;R)$ \cite{To03a,Za09}
and $J_{\nu}(x)$ is the Bessel function of the first kind of order $\nu$.

For typical disordered two-phase media, the variance  $\sigma_{_V}^2(R)$ for large $R$ goes to zero
like $R^{-d}$ \citep{Lu90b,Qu97b} and hence the value of $R$ at which the product $\sigma_{_V}^2(R)\,R^d$
first effectively reaches its asymptote provides a linear measure of the representative elementary volume.
However, for hyperuniform disordered two-phase media, $\sigma_{_V}^2(R)$  goes to zero asymptotically more
rapidly than the inverse of the window volume, i.e., faster than $R^{-d}$, which is equivalent to the following condition
on the spectral density \cite{Za09}:
\begin{eqnarray}
\lim_{|\mathbf{k}|\rightarrow 0}\tilde{\chi}_{_V}(\mathbf{k}) = 0.
\label{hyper-2}
\end{eqnarray}
This hyperuniformity condition dictates that the
direct-space autocovariance function  $\chi_{_V}({\bf r})$ exhibits both positive and negative correlations 
such that its volume integral over all space is exactly zero, i.e., $\int_{\mathbb{R}^d} \chi_{_V}({\bf r}) d{\bf r}=0$,
which is the hyperuniformity sum rule in direct space \cite{To16b}. 
{\it Stealthy hyperuniform} two-phase media  are a subclass of hyperuniform
systems in which $\tilde{\chi}_{_V}(\mathbf{k})$ is zero for a range
of wavevectors around the origin, i.e.,
\begin{equation}
\tilde{\chi}_{_V}(\mathbf{k})= 0 \qquad \mbox{for}\; 0 \le |{\bf k}| \le K,
\label{stealth}
\end{equation}
where $K$ is some positive number. 
All of the models of triply periodic media investigated in this paper are stealthy hyperuniform.

As in the case of hyperuniform point configurations \citep{To03a,Za09,Za11b,To18a},  
when the spectral density has the following power-law form in the limit $|{\bf k}| \to 0$:
\begin{equation}
\tilde{\chi}_{_V}(\mathbf{k})\sim |\mathbf{k}|^{\alpha},
\label{eqn:smallk}
\end{equation}
there are three different scaling regimes (classes) that
describe the associated  large-$R$ behaviors of local
volume-fraction variance \cite{Za09,To18a}:
\begin{eqnarray}  
\sigma^2_{_V}(R) \sim 
\begin{cases}
    R^{-(d+1)}, \quad \qquad \alpha >1 & \mbox{(Class I)}\\
    R^{-(d+1)} \ln R, \quad \alpha = 1 & \mbox{(Class II)},\\
    R^{-(d+\alpha)}, \quad 0 < \alpha < 1 &  \mbox{(Class III)}
\end{cases}
\label{sigma-V-asy}
\end{eqnarray}
where the exponent $\alpha$ is a positive constant.
Class I is the strongest hyperuniformity class, which includes all periodic two-phase media \cite{To18a}, such
as the ones we study in this paper,  as well as certain exotic disordered two-phase media \cite{Fl09b,Fr17,To21a,Ch21,Ki21,To22b,Me23}.
The leading-order asymptotic term in the asymptotic expansion  of $\sigma_{_V}^2(R)$ for class I hyperuniformity, which includes the implied coefficient,
is explicitly given by \cite{Za09,To18a}
\begin{equation}
	\sigma^2_{_V}(R) \sim {\overline B}_V \left(\frac{D}{R}\right)^{d+1},
\end{equation}
where 
\begin{equation}
{\overline B}_V=\frac{\Gamma(1+d/2)}{\pi^{(2+d)/2} D^{d+1} \fn{v_1}{1}}
    \int_{\mathbb{R}^d}  \frac{\spD{{\bf k}}}{k^{d+1}} d{{\bf k}},
\label{Fourier-B}    
\end{equation}
where $D$ is a characteristic ``microscopic" length scale of the medium.
While all class I hyperuniform media have local volume-fraction variances that scale as $R^{-(d+1)}$ for large $R$, the coefficient  ${\overline B}_V$  multiplying $R^{-(d+1)}$ is different among them. Hence,   
${\overline B}_V$  provides a {\it hyperuniformity
order metric} that can be used to rank order different structures according to the degree to which they suppress large-scale local volume-fraction fluctuations \cite{Za09,To18a,To22a}.

\subsection{Pore-Size Functions}
\label{Pore}

We also characterize the pore phase by determining the probability $F(\delta)$ that a randomly placed 
sphere of radius  $\delta$ centered in the pore space ${\cal V}_1$ lies entirely in ${\cal V}_1$. By definition, $F(0)=1$ and $F(\infty)=0$.
The quantity $F(\delta)$ is the complementary cumulative distribution function associated with
the corresponding pore-size probability density function $P(\delta)=-\partial F(\delta)/\partial \delta$.
At the extreme values of $P(\delta)$, we have that $P(0)=s/\phi_1$ and $P(\infty)=0$.
The $n$th moment of the pore-size probability density is defined by \cite{To02a}
\begin{eqnarray}
\langle \delta^n \rangle &\equiv& \int_0^\infty \delta^n P(\delta)\, \dd{\delta} \nonumber\\
&=& n \int_0^\infty \delta^{n-1} F(\delta) \, \dd{\delta}.
\end{eqnarray}
We will be particularly interested in the 
  \textit{mean pore size} $\langle \delta \rangle$ and the \textit{second moment} $\langle \delta^2 \rangle$: 
\begin{align}
  \langle \delta \rangle &= \int_0^{\infty} F(\delta) \, \dd \delta ,
  \label{mean-pore-size}\\
  \langle \delta^2 \rangle &= 2\int_0^{\infty} \delta \, F(\delta) \, \dd{\delta}.
  \label{2nd-moment}
\end{align}
These characteristic length scales of the pore phase have been shown to be related
to certain diffusion properties of the porous medium \cite{To91f} as well as
its fluid permeability \cite{To20}. Note that the pore-size probability function $F(\delta)$ can be easily extracted from 3D digitized images of real porous media \cite{Co96}.

\section{Microstructure-Dependent Formulas to Predict Transport Properties}
\label{transport}

We begin by deriving an expression for the effective electrical conductivity $\sigma_e$
of triply periodic bicontinuous media with $\phi_1=\phi_2=1/2$ using the strong-contrast formalism \cite{To85f,To02a} and then show how this general formula reduces to the exact result for the Schwarz P, the Schwarz D, and the Schoen G minimal surfaces and provides an accurate approximation for other bicontinuous media. 
This derivation is followed by a brief description of known formulas for the fluid permeability and the spreadability that depend on functionals of certain statistical descriptors, which we compute in Sec. \ref{micro}
and then apply in Sec. \ref{estimates} to estimate the transport properties of the five triply periodic
bicontinuous models shown in Fig. \ref{fig:models}.

\subsection{Effective Conductivity and Formation Factor}

The strong-contrast expansions derived by Torquato for the effective conductivity $\sigma_e$ 
of two-phase media in any space dimension $d$ \cite{To85f,To02a}
can be viewed as two different expansions that perturb around the Hashin--Shtrikman
optimal structures. 
As a result, the first few terms of this expansion, beyond the second-order Hashin-Shtrikman terms, should
yield an excellent approximation of $\sigma_e$ for any values of the phase conductivities for dispersions in which
the inclusions are prevented from forming large clusters. In particular, for $d=3$,
its truncation after third-order terms yields the expression \cite{To85f}
\begin{equation}
\sigma_e(\sigma_q,\sigma_p,\phi_p,\zeta_p)=\frac{\sigma_q(1+2\phi_p\beta_{pq}-2\phi_q\zeta_p\beta^2_{pq})}
{1-\phi_p\beta_{pq}-2\phi_q\zeta_p\beta^2_{pq}},
\label{3-point}
\end{equation}
where $p$ and $q$ denote the two different phases 1 or 2 such that $p \neq q$, 
\begin{equation}
\beta_{pq}=\frac{\sigma_p-\sigma_q}{\sigma_p+2\sigma_q}
\end{equation}
and $\zeta_p$ is a three-point microstructural parameter that is a functional of the three-point
correlation function $S_3^{(p)} ({\bf r},{\bf s})$ associated with phase $p$.
Formula (\ref{3-point}) has been shown to provide  highly accurate approximations
of  $\sigma_e$  for a large class of ordered and disordered dispersions in which the particles (phase $p$)
do not form very large clusters \cite{To85f,To02a}.

Importantly, when $\sigma_2 \ge \sigma_1$, $\sigma_e(\sigma_1,\sigma_2,\phi_2,\zeta_2)$, obtained from formula (\ref{3-point})
perturbs about the Hashin-Shtrikman lower-bound structures in which the dispersed phase is the more conducting one relative
to the connected (continuous) matrix phase. However, in the phase interchanged case, i.e.,  $\sigma_e(\sigma_2,\sigma_1,\phi_1,\zeta_1)$ 
obtained from formula (\ref{3-point}),
perturbs about the Hashin-Shtrikman upper-bound structures in which the dispersed phase is  less conducting one relative
to the connected (continuous) matrix phase. 

Consider the mean of these two resulting formulas, i.e.,
\begin{equation}
\sigma_e^* \equiv \frac{\sigma_e(\sigma_1,\sigma_2,\phi_2,\zeta_2)+\sigma_e(\sigma_2,\sigma_1,\phi_1,\zeta_1)}{2}.
\label{3-point-bi}
\end{equation}
Formula (\ref{3-point-bi}) interpolates between the two aforementioned dispersions and topologies
and hence is expected to be a good approximation for a class of bicontinuous media. Indeed,
in the special case $\phi_1=\phi_2=1/2$ and $\zeta_1=\zeta_2=1/2$, formula (\ref{3-point-bi}) yields
\begin{equation}
\sigma_e^*=\frac{\sigma^2_1+4\sigma_1\sigma_2+\sigma^2_2}
{3(\sigma_1+\sigma_2)},
\label{bi-cond}
\end{equation}
which was shown to be exact for the triply periodic bicontinuous composites
separated by the Schwarz P and Schwarz D minimal surfaces \cite{To02d,To03b}.
For these reasons, formula (\ref{3-point-bi}) should also provide accurate estimates of the
effective conductivity for bicontinuous media in the vicinity of $\phi_1=\phi_2=1/2$.

Consider a porous medium whose pore space ${\cal V}_1$ is filled with an electrically conducting fluid of conductivity $\sigma_1$ and a solid phase that is perfectly insulating ($\sigma_2=0$).
The {formation factor} $\cal F$ is defined to be the inverse of the dimensionless effective
conductivity, i.e., 
\begin{equation}
{\cal F}\equiv \sigma_1/\sigma_e.
\label{formation}
\end{equation}
The formation factor $\cal F$ is a measure of the {\it tortuosity} of the
{\it entire} pore space, including topologically connected parts of the
pore space as well as disconnected portions (e.g., isolated pores).
If the pore space does not percolate, then $\cal F$ is unbounded
or, equivalently, $\sigma_e/\sigma_1=0$. Roughly speaking, the formation factor $\cal F$ quantifies the degree of ``windiness" for
electrical transport pathways across a macroscopic sample. 

\subsection{Fluid Permeability}

Avellaneda and Torquato \citep{Av91b} used the solutions of the time-dependent Stokes equations, which can be expressed as a sum of normal modes,
to derive a rigorous relation
 connecting the fluid permeability $\K$ to the formation factor of the porous medium
and a length scale that is determined by the eigenvalues of the Stokes
operator. Specifically, the fluid permeability is exactly given by
\begin{equation}
 \K=\frac{{\cal L}^2}{{\cal F}} \: ,    
\label{k-exact}
\end{equation}
 where $\cal L$  is a certain weighted sum over the {\it viscous relaxation times}
$\Theta_n$ associated with the time-dependent Stokes equations
 (i.e., inversely proportional to the eigenvalues of the Stokes operator).
As noted in the Introduction, the theoretical prediction of the fluid permeability $\K$ for general microstructures is a notoriously difficult problem.
This complexity is due in part to the fact that  $\K$, roughly speaking, may be regarded as an
effective pore channel cross-sectional area of the {\it dynamically} connected
part of the pore space, i.e., the topologically connected portion of the pore space that carries most of the flow, which
eliminates isolated pores and dead-ends as well as connected regions with very little flow \cite{To02a}.  
Various approximations for the permeability $\K$ or length scale $\cal L$  have been put forth
that depend on certain diffusion properties \cite{Jo86,To90e,Av91b}. 

More recently,  cross-property relations \cite{Av91b,To91f} and the exact relation \eqref{k-exact} 
were used to propose the  following approximation for the fluid permeability  \cite{To20}:
\begin{equation}
\K \approx \frac{\langle \delta^2 \rangle}{\cal F}\,,
\label{k-approx}
\end{equation}
where $\E{\delta^2}$ is the second moment of the pore-size probability density function, defined by Eq. (\ref{2nd-moment}).
Note that approximation (\ref{k-approx}) implies  that the exact length scale ${\cal L}$ in (\ref{k-exact}) for the permeability
is approximately given by
\begin{equation}
{\cal L}^2 \approx \langle \delta^2 \rangle.
\end{equation}
It has been shown that the formula (\ref{k-approx}) provides reasonably accurate permeability predictions of nonhyperuniform
and hyperuniform porous media, including periodic media, in which the pore
space is well connected \cite{To20}. We note that $\langle \delta^2 \rangle$ has been recently shown to be related
to the {\it critical pore radius} for certain models consisting of spherical obstacles \cite{Kl21b}.

Rigorous bounds have also been devised that depend on limited microstructural information \cite{Pr61,Be85a,Ru89,To02a}.
In the present work, we will apply the  so-called two-point ``void" upper bound on the fluid permeability of a
general three-dimensional isotropic porous medium \cite{Ru89a}, which is given by 
\begin{equation}
\K \leq \frac{2}{3\phi_2^2} \,\ell^2_P,
\label{k-2pt}
\end{equation}
where $\ell_P$ is the length scale  defined as
\begin{equation}
    \label{lp-1}
    \ell^2_P =
    \int_{0}^{\infty} \chi_{_V}(r)\, r \dd{r},
\end{equation}
where $\chi_{_V}(r)$ is the angular-averaged autocovariance function.
The two-point void bound (\ref{k-2pt}) was originally derived by
Prager \citep{Pr61} with
an incorrect constant prefactor, which was subsequently
corrected by Berryman and Milton \citep{Be85a}
and Rubinstein and Torquato \citep{Ru89a} using different variational
approaches. The two-point void bound (\ref{k-2pt}) on $\K$ has
been generalized to treat two-dimensional
media as well as dimensions higher than three \citep{To02a}.
It is noteworthy that in two dimensions or, equivalently, transversely isotropic media, 
``coated-cylinders" model, which has recently been shown to be hyperuniform  \cite{Ki19a,Ki19b}, realizes the upper bound \eqref{k-2pt}
exactly, implying that this model achieves the maximum permeability
among all microstructures with the
same porosity $\phi_1$ and pore length scale $\ell_P$ \citep{To04a}.

Torquato \cite{To20} obtained a Fourier representation of the three-dimensional length
scale (\ref{lp-1}) in terms of the angular-averaged spectral density $ {\tilde  \chi}_{_V}(k)$, which is given by
\begin{equation}
    \label{lp-2}
    \ell^2_P  =
    \frac{1}{2\pi^2}\int_{0}^{\infty} {\tilde  \chi}_{_V}(k) \dd{k}.
\end{equation}
To compute the two-point void upper bound \eqref{k-2pt} for the model microstructures
considered in this paper, we will be using this Fourier representation of the length scale $\ell_P$.

\subsection{Diffusion Spreadability}

The diffusion spreadability is a dynamical probe that directly links certain time-dependent diffusive transport with the microstructure of heterogeneous media across length scales \cite{To21d}.
Here, one examines the time-dependent problem of mass transfer of a solute in a two-phase medium where all of the solute is initially contained in phase 2, and it is assumed that the solute has the same diffusion coefficient $\cal D$ in each phase. The spreadability $\mathcal{S}(t)$ is defined as the total solute present in phase 1 at time $t$. Torquato demonstrated that the time-dependent diffusion spreadability
$\mathcal{S}(t)$ in any space dimension $d$ is \textit{exactly} related to the microstructure via the autocovariance function $\chi_{_V}(\mathbf{r})$ in direct space or, equivalently, via the spectral density $\tilde{\chi}_{_V}(\mathbf{k})$ in Fourier space \cite{To21d}:
\begin{equation}
 \mathcal{S}(\infty) - \mathcal{S}(t)  =\frac{1}{(2\pi)^d\phi_2}\int_{\mathbb{R}^d} \tilde{\chi}_{_V}(\mathbf{k})\exp(-k^2{\cal D}t) \dd{\mathbf{k}}.
\label{spread}
\end{equation}
Here, $\mathcal{S}(\infty)-\mathcal{S}(t)$ is called the {\it excess spreadability}, where $\mathcal{S}(\infty)=\phi_1$  is the infinite-time limit 
of ${\cal S}(t)$.
The reader is referred to Ref. \cite{To21d} for a description of the remarkable
links between      the spreadability ${\cal S}(t)$, covering problem of discrete geometry,
and nuclear magnetic resonance (NMR) measurements \cite{Mit92b,No14}.

Torquato showed that the
small-, intermediate-, and long-time behaviors of $\mathcal{S}(t)$ are  
directly determined by the small-, intermediate-, and large-scale structural characteristics of the two-phase medium.
Moreover,  when the  spectral densities exhibit the power-law form
(\ref{eqn:smallk}), it was demonstrated that the long-time asymptotic behavior of the excess spreadability 
is given by the inverse power law $1/t^{(d+\alpha)/2}$, implying a faster decay as $\alpha$ increases for some dimension $d$.
Thus, compared to a standard nonhyperuniform medium with a power-law decay $t^{-d/2}$, a hyperuniform medium with a decay rate $t^{-(d+\alpha)/2}$ can be viewed as having an effective dimension that is higher than the space dimension, namely, $d+\alpha$. 
The spreadability has been profitably used
to quantify a myriad of nonhyperuniform and hyperuniform heterogeneous media across length scales \cite{Wa22a,Sk23,Ma23}.

Importantly,  {\it stealthy hyperuniform media}, ordered or not, are characterized by the fastest decay  rates of excess spreadability among
all media with  the infinite-time asymptotes that are approached {\it exponentially}
fast. This latter category includes all of the triply periodic 
bicontinuous media considered in this paper, as explicitly shown
in Sec. \ref{spreadability-estimates}.

\section{Evaluation of Microstructural Descriptors}
\label{micro}

\subsection{Spectral Density}
\label{spec-dens}

Here, we derive explicit formulas for the spectral densities of the triply periodic model microstructures considered in this paper by exploiting the fact that each model can be viewed as a certain packing of identical nonoverlapping inclusions. 
In particular, it follows from Ref. \cite{To16a} that the spectral density
of a general packing (disordered or periodic) packing of oriented, identical nonoverlapping particles, each occupying
region $\Omega$, at number density $\rho$ is given by
\begin{eqnarray}
{\tilde \chi}_{_V}({\bf k})&=& \rho |{\tilde m}({\bf k};\Omega)|^2 S({\bf
k}) \nonumber \\
&=& \phi_1 \frac{|{\tilde m}({\bf k};\Omega)|^2}{|\Omega|} S({\bf k}),
\label{chi-packing-1}
\end{eqnarray}
where  $|\Omega|$ is the volume of an inclusion, $S({\bf k})$ is the structure factor of centroids of the inclusions and ${\tilde m}({\bf
k};\Omega)$ is the Fourier transform of the inclusion indicator function,
which is defined to be
\begin{equation}
m({\bf r};\Omega) =
\begin{cases}
    1, & {\bf r} \in \Omega,    \\
    0, & {\bf
r} \notin \Omega.
\end{cases}
\label{m}
\end{equation}
Here, $\bf r$ is measured with respect to the centroid of the inclusion,
$\phi_1$ is the  fraction of space covered by the nonoverlapping
inclusions, and we note that ${\tilde m}({\bf k=0};\Omega)=|\Omega|$.
Relation (\ref{chi-packing-1})  is a straightforward generalization of the corresponding expression for identical nonoverlapping spheres \cite{To16a}.

Now we note that for any periodic packing of a single inclusion of general shape within a fundamental cell
of volume $v_F$ in $\mathbb{R}^3$, the
structure factor is  given by
\begin{equation}
S({\bf k}) =\frac{(2\pi)^3}{v_F} \sum_{{\bf Q} \neq {\bf 0}} \delta({\bf
k} -{\bf Q}),
\label{factor}
\end{equation}
where the sum is over all reciprocal lattice (Bragg) vectors, except $\bf
Q=0$. Substitution of relation (\ref{factor}) into (\ref{chi-packing-1}) yields
the corresponding spectral density for such a periodic packing:
\begin{eqnarray}
{\tilde \chi}_{_V}({\bf k})&=& \frac{\phi_1(2\pi)^3}{v_F} \frac{|{\tilde
m}({\bf k};\Omega)|^2}{|\Omega|} \sum_{{\bf Q} \neq {\bf 0}}
\delta({\bf k} -{\bf Q}),
\label{chi-packing-2}
\end{eqnarray}
where $\phi_1=|\Omega|/v_F$.

\begin{table*}[htp]
  \caption{Values of the radial spectral densities
  $\tilde{\chi}_{_V}(k)$ at the first four Bragg peaks ($k=Q_n$ for
  $n=1,2,3,4$) for the five
  triply periodic models with porosity $\phi_1=0.5$: Schwarz P, Schwarz
  D, Schoen G, circular-channel model, and square-channel model. We see that the Schwarz $P$ surface
  has the maximum value of the spectral density at the first Bragg peak among all five models.
  Here, we take the side length of the cubic fundamental cell to be unity, i.e., $L=1$.
  \label{tab:peaks}
  }
  \centering
  \begin{tabular}{c|c c c c}
  \hline
  Model & $Q_1 L$ & $Q_2L$& $Q_3L$ &$Q_4L$\\
  \hline
  Schwarz P & $8.3650\times 10^{-2}$ & $9.8353\times 10^{-8}$ &
  $4.8765\times 10^{-3}$ & $7.5304\times 10^{-8}$ \\
  Schwarz D & $8.0497\times 10^{-10}$ & $7.4079\times 10^{-6}$ &
  $3.0979\times 10^{-2}$ & $1.4651\times 10^{-11}$ \\
  Schoen G & $1.2782\times 10^{-5}$ & $4.7427\times 10^{-2}$ &
  $7.0517\times 10^{-9}$ & $2.4953\times 10^{-10}$ \\
  circular-channel & $7.9774\times 10^{-2}$ & $7.2234\times 10^{-5}$ &
  $5.1283\times 10^{-3}$ & $2.4823\times 10^{-4}$ \\
  square-channel & $7.5986\times 10^{-2}$ & $1.8509\times 10^{-6}$ &
  $5.5484\times 10^{-3}$ & $2.8182\times 10^{-6}$ \\
  \hline
  \end{tabular}
  \end{table*}

Now we recognize that for the five triply periodic models considered here,
each pore region within a fundamental cell can be viewed
as a {\it single} concave ``inclusion" with a fixed orientation within a simple cubic lattice of side length $L$ and hence $\phi_1=1/2$. Thus, substituting
$\phi_1=1/2$ and $v_F=L^3$ into Eq. \eqref{chi-packing-2} yields the spectral density for such periodic bicontinuous media to be
\begin{eqnarray}
{\tilde \chi}_{_V}({\bf k})&=& \frac{4\pi^3}{L^3} \frac{|{\tilde m}({\bf
k};\Omega)|^2}{|\Omega|} \sum_{{\bf Q} \neq {\bf 0}} \delta({\bf k}
-{\bf Q}),
\label{chi-packing-3}
\end{eqnarray}
where ${\bf Q}$ represents the reciprocal lattice vectors of the simple cubic lattice. 
Letting $Q_n$ denote the magnitude of the $n$th Bragg peak, the first four Bragg peaks are given by
$Q_1L/(2\pi)=1$,  $Q_2L/(2\pi)=\sqrt{2}$, $Q_3L/(2\pi)=\sqrt{3}$, and $Q_4L/(2\pi)=2$. 
For the applications in this paper, we require the radial spectral density $\spD{k}$, i.e., the {\it angular} average of the directional-dependent
spectral density ${\tilde \chi}_{_V}({\bf k})$ given by Eq. \eqref{chi-packing-3}, yielding
\begin{align}   
    \tilde{\chi}_{_V}(k) = \frac{4\pi^3}{L^3} \sum_{n=1}^\infty Z(Q_n)
\tilde{A}_2(Q_n;\Omega) \frac{\delta(k
-Q_n)}{4\pi {Q_n}^2} ,
\label{chi-packing-4}
\end{align}
where 
\begin{equation}
\tilde{A}_2(Q_n;\Omega) \equiv {Z(Q_n)}^{-1} \sum_{{|\bf Q|} = Q_n} |{\tilde m}({\bf Q};\Omega)|^2  / |\Omega|
\label{A}
\end{equation}
is the angular average of the form factor of the inclusion over all reciprocal lattice vectors whose magnitudes are equal to the $n$th Bragg-peak wavenumbers $Q_n$, $Z(Q_n)$ is the coordination number at radial distance $Q_n$ for a given reciprocal lattice vector, and $\delta(x)$ is a radial Dirac
delta function.

The spectral density formula \eqref{chi-packing-3} is equivalent to the following representation: 
\begin{equation}
\spD{\vect{Q}} = \frac{1}{v_F} \abs{\fn{\tilde{\mathcal{J}}^{(i)}}{\vect{Q}}}^2,
\label{eq:spectral density}
\end{equation}
where $\fn{\tilde{\mathcal{J}}^{(i)}}{\vect{Q}}$ is the Fourier transform of zero-mean indicator
function $\fn{\mathcal{J}^{(i)}}{\vect{r}}$ \cite{To99c}, defined by relation \eqref{J}
 or, equivalently, the Fourier transform of $\fn{m}{\vect{r};\Omega}-\phi_1$ for phase 1, defined in Eq. \eqref{m}.
For the triply periodic bicontinuous models considered in this paper, we evaluate the spectral density via formula (\ref{eq:spectral density}) using efficient fast-Fourier transform (FFT) techniques \cite{Co65,Har20} and then take the angular average for our purposes.

\begin{figure}
\centering
\includegraphics[width=0.4\textwidth]{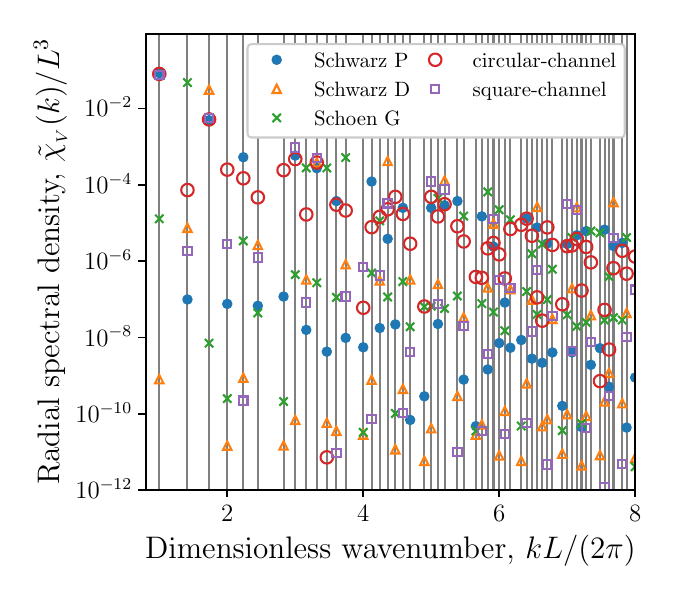}
\caption{Semi-log plot of the radial spectral densities $\tilde{\chi}_{_V}(k)$ as functions of the dimensionless wavenumber $kL/(2\pi)$ for the five triply periodic models with porosity $\phi_1=0.5$: Schwarz P, Schwarz D, Schoen G, circular-channel model, and square-channel model.
Here, $L$ is the side length of the cubic simulation box.
Due to the periodicity of these models, $\spD{k}$ is nonzero only at the Bragg peaks of the simple cubic lattice (shown in gray vertical lines).
\label{fig:chik}
}
\end{figure}

In Fig. \ref{fig:chik}, we plot the radial spectral densities $\tilde{\chi}_{_V}(k)$
as functions of the dimensionless wavenumber $kL/(2\pi)$ for the five
triply periodic models with porosity $\phi_1=0.5$: Schwarz P, Schwarz
D, Schoen G, circular-channel model, and square-channel model.
We also tabulate the corresponding values of $\spD{k}$ at the first four Bragg peaks $k=Q_n$ for $n=1,2,3,4$; see Table \ref{tab:peaks}.

\subsection{Local Volume-Fraction Variance}

Substitution of Eq. \eqref{chi-packing-3} for the spectral density into Eq. \eqref{phi-var-2} yields an explicit expression for the local volume-fraction variance $\sigma_{_V}^2(R)$ for the triply periodic media considered here:
\begin{align}
\sigma_{_V}^2(R) &= \frac{1}{2 \pi^2 v_1(R)} \int_0^\infty k^2{\tilde
\chi}_{_V}(k) {\tilde \alpha}_2(k; R) \dd{k} \nonumber \\
  &= \frac{1}{2 v_1(R) L^3}
\sum_{n=1}^\infty Z(Q_n) \tilde{A}_2(Q_n;\Omega)   {\tilde \alpha}_2(Q_n;
R)
\nonumber \\
&=
\frac{3\pi^2}{v_1(R) L^3}
\sum_{n=1}^\infty Z(Q_n) \tilde{A}_2(Q_n;\Omega)
\frac{J^2_{3/2}(Q_nR)}{{Q_n}^3},
\label{phi-var-3}
\end{align}
where we have used the angular-averaged formula (\ref{chi-packing-4})
for the spectral density. Substitution of Eq. \eqref{chi-packing-4} into Eq.
\eqref{Fourier-B} yields the corresponding expression for the large-$R$ asymptotic coefficient \cite{Ki21}:
\begin{align}
     \overline{B}_V =
     \frac{9}{4} \sum_{n=1}^\infty Z(Q_n)
\frac{\tilde{A}_2(Q_n;\Omega)}{L^3 (Q_nL)^4}. \label{eq:Bv-periodic}
\end{align}

\begin{figure}
\centering
\includegraphics[width=0.4\textwidth]{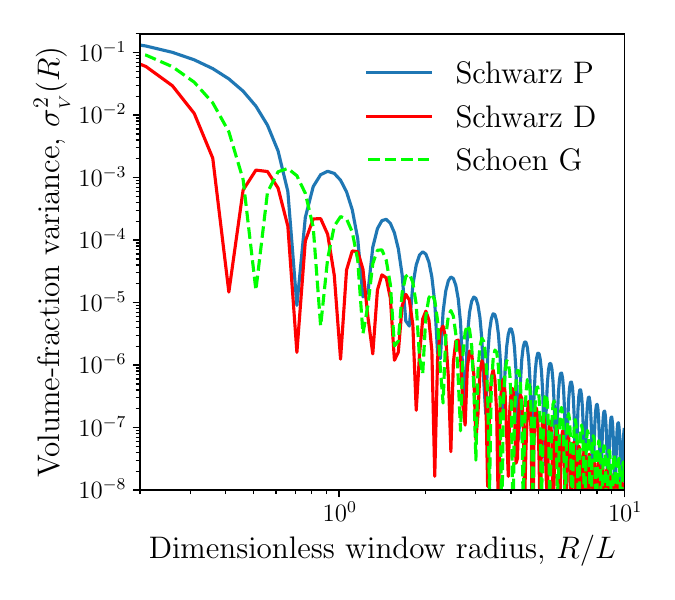}
\caption{Log-log plot of the local volume-fraction variance
$\sigma_{_V}^2(R)$
as a function of the dimensionless window radius $R/L$ for the three
triply periodic minimal surfaces with porosity
$\phi=0.5$: Schwarz P, Schwarz D, and Schoen G.
For large radii ($R/L \gg 1$), the variance $\sigma_{_V}^2(R)$ of all
models commonly decays as fast as $R^{-4}$. On average, the Schwarz P and
D structures have the
largest and smallest variances, respectively.
\label{fig:variance}
}
\end{figure}

We compute the local volume-fraction variance $\sigma_{_V}^2(R)$ as a
function of window radius $R$ for the five triply periodic structures
with $\phi_1=1/2$ from Eq. \eqref{phi-var-3}. 
Since these five structures are periodic and, hence,  class I hyperuniform,
their variances decay as fast as $R^{-4}$ for large radii ($R\gg L$).
In Fig. \ref{fig:variance}, we plot the variances on a log-log scale for the three triply periodic minimal surfaces: Schwarz P, Schwarz D, and Schoen G.
The results for the square- and circular-channel models are not shown, since they are virtually indistinguishable from that of the Schwarz P on the scale of this figure.
The values of the large-$R$ asymptotic coefficient $\overline{B}_V$
 for the five triply periodic structures, as computed from Eq.
\eqref{eq:Bv-periodic} and listed in Table \ref{tab:Bv}, enable us to rank order the structures according to their large-scale volume-fraction fluctuations. 
Table \ref{tab:Bv} also includes the result for a spherical obstacle at the centroid of the cubic cell (spherical-obstacle model) at $\phi_1=1/2$.
We see that the Schwarz P structure has the largest large-scale
volume-fraction fluctuations, followed by the circular-channel model,
the square-channel model, the spherical-obstacle model, the Schoen G structure, and
finally, the Schwarz D structure has the smallest value of $\overline{B}_V$.
Importantly, we will see in Sec. \ref{permeability-estimates} that these 
rankings of the structures are wholly consistent with the rankings of their fluid
permeabilities.

\begin{table}
\caption{
Hyperuniformity order metric $\overline{B}_V$ for six triply
periodic models with porosity $\phi=1/2$: Schwarz P, Schwarz D,
Schoen G, circular-channel model, square-channel model, and
spherical-obstacle model.
The models are arranged in an ascending order of $\overline{B}_V$.
The quantities are computed by taking the side length of the cubic fundamental cell
to be unity, i.e., $L=1$.
\label{tab:Bv}
}
\centering
\begin{tabular}{c|c }
\hline
Model & $\overline{B}_V$ \\
\hline
Schwarz P & $4.936 \times 10^{-4}$ \\
circular-channel & $4.721 \times 10^{-4}$ \\
square-channel & $4.504 \times 10^{-4}$ \\
spherical obstacle & $4.462 \times 10^{-4}$  \\
Schoen G & $1.376 \times 10^{-4}$  \\
Schwarz D & $6.002 \times 10^{-5}$ \\
\hline
\end{tabular}
\end{table}

\begin{figure}[h]
  \centering
  \includegraphics[width=0.4\textwidth]{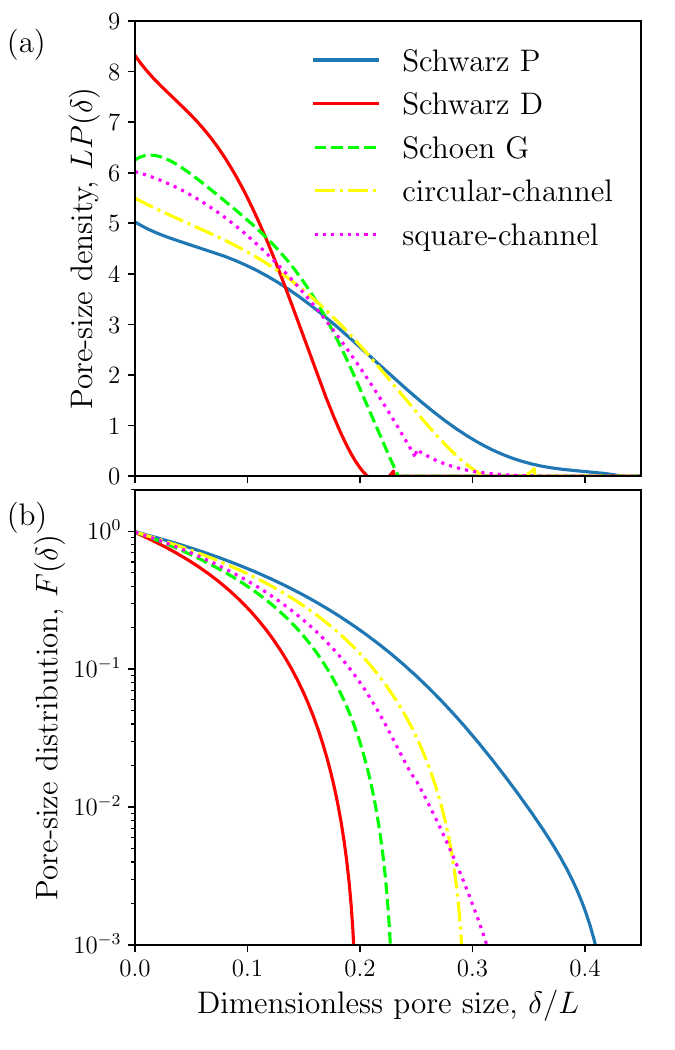}
  \caption{Plots of (a) the dimensionless pore-size probability density function $L P(\delta)$ and (b)
  the corresponding complementary cumulative distribution function $F(\delta)$ as functions of dimensionless pore radius $\delta/L$ for the five triply periodic models
  with porosity $\phi_1 = 0.5$: Schwarz P, Schwarz D, Schoen G,
  circular-channel model, and square-channel model. Here, $L$ is the
  side length of the cubic periodic simulation box. The curves, which are in very good agreement
  with direct simulations, are generated from the polynomial fits  \eqref{eq:P_Poresize}-\eqref{eq:square_Poresize}.
  \label{fig:pore-size}
  }
  \end{figure}    

\subsection{Pore-Size Functions}

To compute rigorous bounds on the mean survival time, Gevertz and Torquato \cite{Ge09a} obtained 
accurate quintic polynomial fits of pore-size density function $P(\delta)$ for four
triply periodic models: the circular-channel model, and Schwarz P,
Schoen G, and Schwarz D structures.
Here, we improve these polynomial expressions for $P(\delta)$ by
imposing the exact constraint at the origin, i.e., $P(0)=s/\phi_1$, from
independent direct numerical simulations of the pore-size density function.
Furthermore, via an additional simulation, we obtain an accurate
 polynomial fit of $P(\delta)$ for the
square-channel model, which exactly dictates a cubic polynomial for this model.
To summarize, we find the following accurate fits of $P(x)$ for 
all five triply periodic bicontinuous models, where $x\equiv \delta /L$:

\begin{align}
    &\text{Schwarz P:}~ P(x) \nonumber \\
    &=
    \begin{cases}
        \parbox{0.28\textwidth}{$-3443x^5+3807.4x^4-1336.66x^3$ \\  $~~~+149.273x^2-13.7109x+5.02661$,}
    & x < 0.441,
    \\ 0, &\text{otherwise},
            \end{cases} \label{eq:P_Poresize}
            \\
    &\text{Schwarz D:}~ P(x) \nonumber \\
    &=
    \begin{cases}
        \parbox{0.28\textwidth}{
            $9760x^5+8302.1x^4-4002.49x^3$ \\
            $~~~+347.050x^2-32.1064x+8.31791$, 
        }
        & x < 0.230 \\ 
        0, &\text{otherwise},
        \end{cases} 
        \label{eq:D_Poresize}
            \\
    &\text{Schoen G:}~ P(x) \nonumber \\
    &=
        \begin{cases}
            \parbox{0.28\textwidth}{    $55486x^5-33537.6x^4+6997.17x^3$ \\
            $~~~-691.054x^2+15.2274x+6.25259$,} 
        & x <  0.238 \\ 
            0, &\text{otherwise},
        \end{cases}
        \label{eq:G_Poresize}
        \\
            &\text{circular-channel:}~ P(x) \nonumber \\
    &=
        \begin{cases}
        \parbox{0.28\textwidth}{
            $2935x^5-813.9x^4-214.01x^3$\\
            $~~~+37.925x^2-11.7534x+5.49471,$}
         & x < 0.355 \\
        0, &\text{otherwise},
        \end{cases} 
        \label{eq:circular_Poresize}
        \end{align}
    \begin{align}
    &\text{square-channel:}~ P(x) \nonumber \\
    &=  \begin{cases}
            -66.9x^2-5.98x+6.021, & x < 0.25 \\
    -534x^3+548.2x^2-188.66x+21.775, & 0.25 < x < 0.352 \\
    0, &\text{otherwise}
            \end{cases}.
        \label{eq:square_Poresize}
    \end{align}

    \begin{table}[h!]
      \caption{The dimensionless first moment $\langle \delta\rangle/L$
(mean pore size) and second moment $\langle \delta^2\rangle/L^2$ of
the pore-size density function $P(\delta)$ for the five triply periodic
models at porosity $\phi_1=0.5$, as obtained from the polynomial fits
given in Eqs.
\eqref{eq:P_Poresize}-\eqref{eq:square_Poresize}, which are in excellent
agreement
with direct simulations of these moments. We also list the corresponding values of the inverse of the
dimensionless specific surface $(sL)^{-1}$ taken from Ref. \cite{Ju05}.
Models are arranged in an
ascending order of $\langle \delta^2\rangle/L^2$ from bottom to top.
Note that the same ordering applies to the mean pore size $\langle
\delta\rangle/L$  and $(sL)^{-1}$.
      Here, $L$ is the side length of the cubic unit cell.
      }
      \label{tab:moments}
      \centering
{\scriptsize
       \begin{tabular}{c|c c c}
           \hline
           Model    & $\langle \delta\rangle/L$
& $\langle \delta^2\rangle/L^2$  & $(sL)^{-1}$      \\
           \hline
           Schwarz P&
          1.222(8)$\times10^{-1}$&    2.218(1)$\times10^{-2}$ & $4.219
\times 10^{-1}$ \\ 
           circular-channel&
          1.080(7)$\times10^{-1}$&
          1.681(3)$\times10^{-2}$ & $3.788 \times 10^{-1}$ \\ 
           square-channel&
          9.726(1)$\times10^{-2}$&
          1.379(4)$\times10^{-2}$ & $3.333 \times
10^{-1}$ \\
           Schoen G&
          8.812(1)$\times10^{-2}$&
          1.104(1)$\times10^{-2}$ & $3.197 \times 10^{-1}$ \\ 
           Schwarz D&
          7.104(5)$\times10^{-2}$&    7.274(4)$\times10^{-3}$ & $2.563
\times 10^{-1}$ \\
           \hline
   \end{tabular}
}
\end{table}

Figure \ref{fig:pore-size} depicts the dimensionless pore-size probability
density function $L P(\delta)$ and
the corresponding complementary cumulative distribution function
$F(\delta)$ as functions of
dimensionless pore radius $\delta/L$ for the five triply
periodic models with $\phi_1=1/2$, as computed from the polynomial fits
\eqref{eq:P_Poresize}-\eqref{eq:square_Poresize}.
As expected, both functions for  each model monotonically decrease with
$\delta/L$
within a compactly supported interval, i.e., the functions are non-zero up
to a finite and positive value of $\delta/L$.
The Schwarz P  structure has the largest such support among all structures, since it is capable of
accommodating the largest spherical region at the centroid. By contrast, the Schwarz
D structure has the smallest support among all structures, since  it is characterized by
narrow channels that cannot support a large pore, as can be inferred visually from 
the corresponding image shown in Fig. \ref{fig:models}.
The support sizes of the circular-channel and square-channel models
are similar and lie between those of the Schwarz P and Schoen G structures.

The  values of the dimensionless mean pore size or first moment $\langle \delta
\rangle/L$ and second moment $\langle
\delta^2\rangle/L^2$ of the five models computed from the fit functions are listed in Table \ref{tab:moments}. These estimates are in excellent agreement with corresponding simulation results with relative errors no larger than one percent. 
The table also includes the corresponding
values of the inverse of the dimensionless
specific surface $(sL)^{-1}$ taken from Ref. \cite{Ju05}.
The relative rankings of the structures according to both the mean pore
size and second moment as well as  the inverse of the specific
surface are the same: the values of the three quantities are
largest for the Schwarz P structure, followed by the
circular-channel model,
square-channel model, 
Schoen G structure, and Schwarz D structure, which has the smallest ones.

\section{Microstructure-Dependent Estimates of Transport Properties}
\label{estimates}

\subsection{Diffusion Spreadability}
\label{spreadability-estimates}

For the models of periodic media considered in this paper, we can obtain
an explicit formula for the diffusion spreadability by substituting relation (\ref{chi-packing-3})
for the spectral density into the general expression (\ref{spread}) for the spreadability (with $\phi_1=1/2$)
to yield
\begin{align}
    \mathcal{S}(\infty) - \mathcal{S}(t)
    =& \frac{1}{\pi^2}\int_0^\infty k^2 \tilde{\chi}_{_V}(k)\exp(-k^2{\cal D}t) \dd{k} \nonumber\\
    =&
    \frac{1}{L^3} \sum_{n=1}^\infty Z(Q_n) \tilde{A}_2(Q_n;\Omega)
\exp(-{Q_n}^2 \mathcal{D}t)  \label{eq:S_exc2},
\end{align}
where we have used the angular-averaged formula (\ref{chi-packing-4}) for the spectral density
with $\tilde{A}_2(Q_n;\Omega)$ given by (\ref{A}).
For the same reasons noted in Ref. \cite{To21d}, the first term in the sum of
Eq. \eqref{eq:S_exc2} is the dominant contribution  at long times $(\mathcal{D}t/L^2 \gg 1)$ and so
all decay with the same exponential rate given by
\begin{align}   \label{eq:S_exc3}
    &\mathcal{S}(\infty) - \mathcal{S}(t)
    \sim C_{\cal S}(\Omega)\exp(-(2\pi)^2 \mathcal{D}t/L^2) \quad
(\mathcal{D}t/L^2 \gg 1),
\end{align}
where $C_{\cal S}(\Omega)$ is a structure-dependent constant given by
\begin{equation}
C_{\cal S}(\Omega)=\frac{6 \tilde{A}_2(2\pi/L;\Omega)}{L^3},
\label{C}
\end{equation}
and we have used the fact that the first Bragg peak is $Q_1=2\pi/L$ and $Z(Q_1)=6$ for the simple cubic lattice.

\begin{figure}
\centering
\includegraphics[width=0.4\textwidth]{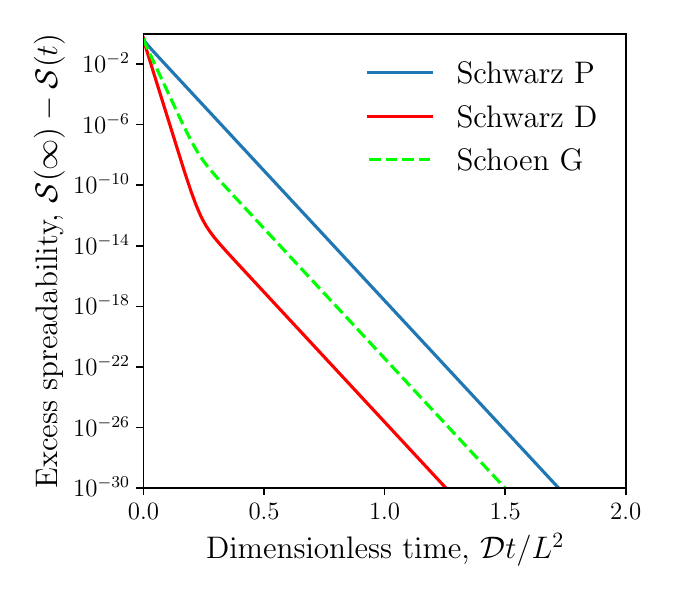}
\caption{
Semi-log plot of the excess spreadability $\mathcal{S}(\infty) -
\mathcal{S}(t)$ as a function of dimensionless time $\mathcal{D}t/L^2$
for the three triply periodic minimal surfaces with porosity
$\phi_1=1/2$: Schwarz P, Schwarz D, and Schoen G.
The Schwarz P and D structures  have the largest and smallest
excess spreadabilities $\mathcal{S}(\infty) - \mathcal{S}(t)$, respectively. Curves for the circular- and square-channel models are not shown here
because they are only slightly lower than but virtually
identical to Schwarz P on the scale of this figure.
\label{fig:spread}
}
\end{figure}

Figure  \ref{fig:spread} shows the excess spreadability $\mathcal{S}(\infty) 
-\mathcal{S}(t)$ as a function of the dimensionless time $\mathcal{D}t/L^2$, as obtained from Eq. \eqref{eq:S_exc2}, for the Schwarz P, Schwarz D, and Schoen G structures. 
The corresponding curves for the circular- and square-channel models are indistinguishable from Schwarz P on the scale of this plot and so are not shown. 
We find that for a specific time, the excess spreadability is the largest for the Schwarz P structure, followed by the circular-channel model, square-channel models, Schoen G, and finally, the Schwarz D has the smallest excess spreadability. 
Importantly, these rankings are consistent with those of the fluid permeabilities, as shown below, as well as other structural descriptors, as described in Sec. \ref{discuss}.
It is noteworthy that the asymptotic formula \eqref{eq:S_exc3} is virtually identical to the exact relation \eqref{eq:S_exc2} (relative errors within the $10^{-7}$ percent) for $\mathcal{D}t/L^2>1$, which is consistent with the values of  the asymptotic coefficient $C_{\cal S}$
listed in Table. \ref{tab:constant}.

\begin{table}[h]
    \caption{The long-time asymptotic
coefficient $C_{\cal S}$ of the excess spreadability $\mathcal{S}(\infty) -
\mathcal{S}(t)$, defined by  Eq. \ref{C}, for the
five triply periodic structures with porosity $\phi_1=1/2$ studied in
this paper. }
    \label{tab:constant}
    \centering
    \begin{tabular}{c|c}
    \hline
    Models  & Coefficient $C_\mathcal{S}$   \\
    \hline
    Schwarz P           & $3.346 \times 10^{-1}$\\
    circular-channel    & $3.191 \times 10^{-1}$\\
    square-channel      & $3.039 \times 10^{-1}$\\
    Schoen G            & $5.113 \times 10^{-5}$\\
    Schwarz D           & $3.220 \times 10^{-9}$\\
    \hline
    \end{tabular}
\end{table}

\subsection{Fluid Permeability}
\label{permeability-estimates}

Here we estimate the fluid permeabilities of the five triply periodic bicontinuous models using the approximation formula (\ref{k-approx}) and two-point void upper bound \eqref{k-2pt}.
Note that the formation factor in Eq. \eqref{k-approx} for the triply periodic minimal surfaces is exactly given 
\begin{equation}
{\cal F}=3,
\end{equation}
where we have used the exact expression (\ref{bi-cond}) with $\sigma_2=0$.
Moreover, this relation is an excellent approximation for the two pore-channel models, since $\zeta_2$ will be very insensitive to the small differences in geometries from the Schwarz P structure and much more sensitive to their topological similarities.
To compute $\K$ from Eq. \eqref{k-approx}, in addition to utilizing ${\cal F}=3$, we employ the second moments $\langle{\delta^2}\rangle$
given in Table \ref{tab:moments}.

\begin{table*}[h!]
    \caption{Estimates of the dimensionless fluid permeability $\K/L^2$ for the five triply periodic models at porosity $\phi_1=1/2$ considered in this paper.
    For each model, the values are estimated via the direct
simulations in Ref. \cite{Ju05}, the
formula \eqref{k-approx}, and  two-point void bound \eqref{k-2pt-Z3}.
    In each column, the value in the square brackets represents the ratio of the estimates of the corresponding model to that of Schwarz P.  Models are arranged in ascending order of $\K/L^2$ from the bottom to the top.
    Here, $L$ is the side length of the cubic fundamental cell.  
     }
    \label{tab:permeability}
    \centering
    \begin{tabular}{c|cc | cc |cc}
    \hline
    Model & \multicolumn{2}{c|}{Direct simulation} &
\multicolumn{2}{c|}{Approximation: \eqref{k-approx}} &
\multicolumn{2}{c}{Two-point bound: Eq. \eqref{k-2pt-Z3}} \\
   \hline
Schwarz P& $3.4765 \times 10^{-3}$ & [ $1.000$ ] & $7.4211 \times
10^{-3}$ & [ $1.000$ ] & $1.2197 \times 10^{-2}$ & [ $1.000$ ] \\
circular-channel& $3.4596 \times 10^{-3}$ & [ $0.995$ ] & $5.5957
\times 10^{-3}$ & [ $0.754$ ] & $1.1741 \times 10^{-2}$ & [ $0.963$ ]
\\
square-channel& $3.0744 \times 10^{-3}$ & [ $0.884$ ] & $4.6051 \times
10^{-3}$ & [ $0.621$ ] & $1.1258 \times 10^{-2}$ & [ $0.923$ ] \\
Schoen G& $2.2889 \times 10^{-3}$ & [ $0.658$ ] & $3.6824 \times
10^{-3}$ & [ $0.496$ ] & $6.5874 \times 10^{-3}$ & [ $0.540$ ] \\
Schwarz D& $1.4397 \times 10^{-3}$ & [ $0.414$ ] & $2.4244 \times
10^{-3}$ & [ $0.327$ ] & $4.3327 \times 10^{-3}$ & [ $0.355$ ] \\
\hline
\end{tabular}
\end{table*}

The  length scale $\ell_P$
involved in the two-point void upper bound
\eqref{k-2pt} on  $\K$ can be expressed explicitly for these models by substituting Eq. \eqref{chi-packing-4} into Eq. \eqref{lp-2} 
\begin{align}
    \ell_P^2 =& \frac{1}{2L^3}
    \sum_{n=1}^\infty Z(Q_n)
    \frac{\tilde{A}_2(Q_n;\Omega)}{{Q_n}^2} \label{lp-Z3},
\end{align}
and, thus, the upper bound on the permeability can be expressed as
\begin{align}
\K \leq & \frac{1}{3{\phi_2}^2 L^3}
    \sum_{n=1}^\infty Z(Q_n)
    \frac{\tilde{A}_2(Q_n;\Omega)}{{Q_n}^2} \label{k-2pt-Z3}.
\end{align}

In  Table \ref{tab:permeability}, we compare the estimates of the dimensionless fluid permeability $\K/L^2$ from two
microstructure-dependent formulas as obtained from  Eqs. \eqref{k-approx} and
\eqref{k-2pt-Z3} to direct simulations  via the immersed-boundary
finite-volume method \cite{Ju05}. The formula \eqref{k-approx} yields estimates
that are within
about a factor of two of the simulation results, which is actually
relatively accurate for a permeability approximation that can applied 
to a broad range of structures; see Refs. \cite{Mar94} and \cite{To02a} and references therein.
While the void upper bound is not sharp, it provides the correct rankings of the 
structures according to their permeabilities, as does the approximation \eqref{k-approx}.
The numbers indicated within the square brackets
in each column of the table represent the values of $\K/L^2$ scaled by that of Schwarz P estimated by the associated method.
Note that all three scaled estimates yield identical relative rankings of $\K/L^2$ for the five models: Schwarz P has the largest value, followed by 
the circular-channel model, the square-channel model, Schoen G, and finally, the Schwarz D has the lowest value.
Thus, the estimate \eqref{k-approx} and \eqref{k-2pt-Z3} can be employed to estimate $\K/L^2$ given the
corresponding reference
value for the Schwarz P structure.
These rankings of $\K/L^2$ are consistent with the rankings of $\E{\delta^2}$, demonstrating  that the structures that have
larger pores on average tend to have higher permeabilities.

\section{Generalized Maximum-Permeability Conjecture for All Porosities}
\label{max-perm}

The maximum-permeability conjecture (see {\bf Conjecture \ref{conj1}} of the Introduction)
applies to the special porosity value of $1/2$. Here we propose the following
generalization of {\bf Conjecture \ref{conj1}} that applies for all nontrivial porosities, i.e.,
$0 < \phi_1 <1$:
\smallskip

\begin{conj}\label{conj2}
    Among three-dimensional porous media at some porosity $\phi_1 \in (0,1)$ within a simple cubic fundamental cell of side length $L$ under periodic boundary conditions, the dimensionless isotropic fluid permeability $\K/L^2$ is maximized for the medium in which the pore space is simply connected with an interface that minimizes the total interface area.
\end{conj}
This conjecture is based on the fact that simply connected pore spaces with smaller total interfacial areas should offer less resistance to Stokes flow (due to the non-slip boundary condition) and, hence, larger fluid permeabilities.
Here we provide rigorous theoretical arguments that support {\bf Conjecture \ref{conj2}} at the two extreme limits of the porosity, i.e., $\phi_1 \to 0$ and $\phi_1 \to 1$. 
First,  we describe what geometries, at these two limits, minimize the total surface area $\cal A$ within a cubic box of side length $L$, i.e., minimizes the specific surface $s={\cal A}/L^3$, without regard to fluid transport.
It is well-known that simple, single convex shapes minimize $\cal A$, depending on the
fraction $\phi$ of space occupied by these domains; they are the sphere, the circular cylinder aligned with a cube edge
and the square slab whose planar interfaces are parallel to a cube face \cite{St16}. Among these three geometries, the sphere has the smallest specific surface in the volume-fraction interval $[0,4\pi/81\approx 0.1551)$, the cylinder has the smallest specific surface in the volume-fraction interval $[4\pi/81,\pi^{-1}\approx 0.3183)$, and the square slab has the smallest specific surface
for $\phi > \pi^{-1}$. 

\begin{figure}
\centering
\includegraphics[width=0.4\textwidth]{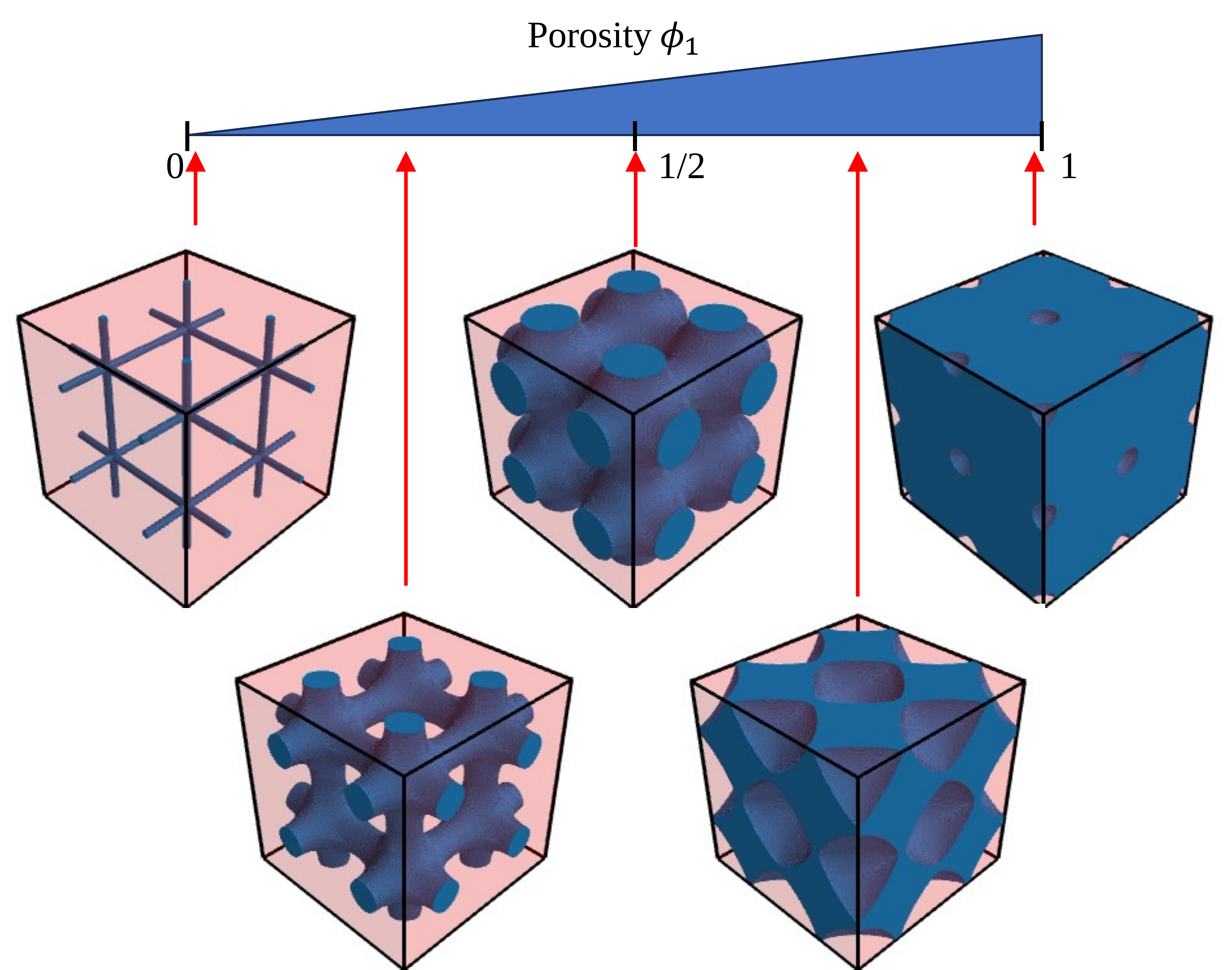}
\caption{The top row of configurations shows the conjectured  Schwarz P structure that maximizes
the fluid permeability $\K/L^2$ as well as the triply periodic models that we have shown maximize $\K/L^2$ 
in the  limit $\phi_1\to 0$ (circular-channel model) and limit $\phi_1\to 1$  (spherical obstacle).
 The bottom row of images, from left to right, show schematics
of the triply periodic structures that are expected to maximize the fluid permeability at porosities
that lie in between 0 and $1/2$ and in between 1/2 and $1$, respectively. The solid phase in the left panel of the bottom row is no longer
connected and remains disconnected until $\phi_1$ approaches unity when it becomes a sphere. In all cases, $2\times 2 \times 2$ fundamentals cells are presented,
where opaque blue and transparent red depict the
domains of the porous phase and solid phase, respectively.
\label{5-vol-frac}
}
\end{figure}

Now consider the geometry that maximizes the fluid permeability in the low-porosity limit, i.e., $\phi_1 \to 0$.
In this case, it is clear that the pore space must be simply and topologically connected in all three principal directions. This condition eliminates a very small fluid-filled sphere, which has a zero permeability. Thus, given that a cylinder has the smallest specific surface among all simply connected pore regions at low porosities, it is natural to consider a pore space consisting of circular cylinders that leads to an isotropic fluid permeability, namely, the circular-channel model consisting of three such tubes that are aligned along the principal axes and that intersect at the centroid of the cubic cell.
The exact solution for such a geometry can easily be obtained from the well-known solution for a single circular cylindrical tube of radius $a$ and length $L$ that passes through the centroid of the fundamental cell and aligned with a cube edge \cite{Sc74,To02a}, which is given by
\begin{align}    \label{eq:K-single-tube}
\K= \frac{a^2}{8}\phi  
\end{align}
and is known to maximize the fluid permeability in a single direction.
Given that $\phi=\pi(a/L)^2$, we can rewrite Eq. \eqref{eq:K-single-tube} in terms of the length scale $L$:
\begin{equation}
\K=\frac{L^2}{8\pi} \phi^2.
\end{equation}
From this solution and the fact that the total porosity of the circular-channel model is $\phi_1= 3\phi$ in the limit $\phi_1\to 0$, we arrive at the exact  expression for the aforementioned three intersecting cylindrical tubes (see the leftmost structure
in the top row of Fig.  \ref{5-vol-frac}) that maximizes the isotropic
fluid permeability  in the low-porosity limit:
\begin{equation}
\K \approx\frac{L^2}{72\pi} \phi_1^2 \qquad (\phi_1 \to 0).
\label{k-tube}
\end{equation}

Next, consider the geometry that maximizes the fluid permeability in the high-porosity limit, i.e., $\phi_1 \to 1$.
In this limit, the geometry must involve Stokes flow around a single solid body, which must be spherical
in order to achieve an isotropic permeability (see rightmost structure
in the top row of Fig.  \ref{5-vol-frac}).  It is well-known that such a body minimizes the drag
force \cite{Hap83} or, equivalently, maximizes the permeability, which is inversely related to the drag force \cite{To02a},
which in the dilute-particle-concentration limit is given by 
\begin{align}
\K =&  \frac{2a^2}{9 (1-\phi_1)}    \qquad (\phi_1 \to 1),
\end{align}
where we have used $1-\phi_1=4\pi(a/L)^3/3$ to obtain the last line.
Given that $1-\phi_1=4\pi(a/L)^3/3$, we can rewrite this expression in terms of the length scale $L$:
\begin{equation}
\K=\frac{L^2}{18(1-\phi_1)^{1/3}}\left(\frac{6}{\pi}\right)^{2/3} \qquad (\phi_1 \to 1).
\end{equation}
As shown above, a sphere at low porosities
has the smallest specific surface measure with respect to the cube side length $L$. This is consistent with the fact that a spherical obstacle
maximizes the isotropic fluid permeability $\K$ in the limit $\phi_1\to 1$.

For porosities intermediate between 0 and $1/2$, the latter at which the optimal structure is conjectured to be the Schwarz P medium \cite{Ju05} (middle image in the top
row of  Fig.  \ref{5-vol-frac}), the maximum-permeability structures must still be bicontinuous with a pore space
that is topologically equivalent (i.e., {\it homeomorphic})
to that of the Schwarz P structure, i.e., orthogonally oriented cylinder-like channels that intersect at the centroid of the cube.
As $\phi_1$ gradually increases from $0$ to $1/2$, the infinite curvature at 
the cylinder junctures  must become finite, leading to increasingly smoother juncture regions
in cylinder-like channel geometries as $\phi_1$ increases until the very smooth Schwarz P minimal surface is reached at $\phi_1=1/2$. 
The leftmost image in the bottom row of Fig.  \ref{5-vol-frac} schematically shows such an optimal structure with a porosity  that lies in between 0 and $1/2$
and that putatively has a minimal interfacial area. As $\phi_1$ gradually increases beyond $\phi_1=1/2$, the Schwarz P surface deforms continuously
(with a concomitant increase in the  cross-sectional area of cylinder-like channels) such that the connected solid phase eventually has a ``pinch-off" porosity point 
where the solid phase becomes disconnected
 until $\phi_1$ approaches unity when it becomes an isolated sphere. 
 The rightmost image in the bottom row of Fig.  \ref{5-vol-frac} schematically shows such an optimal structure with a porosity that lies in between $1/2$ and $1$ and that putatively has a minimal interfacial area. 
 We note that a family of surfaces possessing the symmetry and topology of the Schwarz P surface but with different porosities were presented in Ref. \cite{Ju97}, but these structures were not shown to possess minimal specific surfaces.

\section{Discussion}
\label{discuss}

We have determined various microstructural characteristics, including the spectral densities, pore-size distribution functions,
 local volume-fraction variances and hyperuniformity order metrics of the five triply periodic bicontinuous media models
shown in  Fig. \ref{fig:models}:  Schwarz P,  Schwarz D, and  Schoen G minimal structures as well as the circular-channel and square-channel models.  We also computed the formation factor $\cal F$
and combined this information with the second moment of the pore-size function
to estimate,  for the first time, the fluid permeability $\K$ for all five models via the explicit microstructure-dependent formula (\ref{k-approx}).
These predictions are shown to be in relatively good agreement with direct computer
simulations of the permeabilities \cite{Ju05}.
While the void bound (\ref{k-2pt}), which we also computed via the spectral density, lay appreciably above the simulation data, it provides the same trends in the relative rankings of the five models \st{in} which, consistent with {\bf Conjecture \ref{conj1}}, identifies the Schwarz P structure with the highest permeability. 
We also calculated the time-dependent diffusion spreadability ${\cal S}(t)$ as a function of time $t$ for all five models, which again shows the same trends as for $\K$.

\begin{table*}[bthp]
    \caption{Summary of the key structural and physical properties for the
five triply periodic structures with porosity $\phi_1=1/2$ studied in
in this paper: inverse of the dimensionless specific surface
$(sL)^{-1}$ (taken from Ref. \cite{Ju05}), hyperuniformity order
metric $\overline{B}_V$ (Table \ref{tab:Bv}), the dimensionless pose-size second
moment  $\langle\delta^2 \rangle/L^2$ (Table
\ref{tab:moments}), dimensionless fluid permeability $\K/L^2$ (the
second column of Table \ref{tab:permeability}), and the long-time
asymptotic coefficient $C_\mathcal{S}$ of the excess spreadability
$\mathcal{S}(\infty) - \mathcal{S}(t)$ [see Eq. \eqref{eq:S_exc3}].
    Here, $L$ is the side length of the cubic fundamental cell.
    \label{tab:summary}}
    \centering
    \begin{tabular}{c|c  c  c  c  c}
    \hline
    Quantity & Schwarz P & circular-channel & square-channel & Schoen
G & Schwarz D \\
    \hline
    $(sL)^{-1}$
    & $4.219 \times 10^{-1}$ &$3.788 \times 10^{-1}$ &$3.333 \times
10^{-1}$ &$3.197 \times 10^{-1}$ &$2.563 \times 10^{-1}$ \\
    $\overline{B}_V$  & $4.936 \times 10^{-4}$ & $4.721 \times
10^{-4}$ & $4.504 \times 10^{-4}$ & $1.376 \times 10^{-4}$  & $6.002
\times 10^{-5}$ \\
    $\langle \delta^2 \rangle/L^2$ & 2.218$\times10^{-2}$ &
    1.681$\times10^{-2}$ &
    1.379$\times10^{-2}$ &
    1.104$\times10^{-2}$ &
    7.274$\times10^{-3}$ \\
    $\K / L^2$ &
    $3.4765 \times 10^{-3}$ &
    $3.4596 \times 10^{-3}$ &
    $3.0744 \times 10^{-3}$ &
    $2.2889 \times 10^{-3}$ &
    $1.4397 \times 10^{-3}$ \\
    $C_\mathcal{S}$ &
    $3.346 \times 10^{-1}$& $3.191 \times 10^{-1}$& $3.039 \times
10^{-1}$& $5.113 \times 10^{-5}$& $3.220 \times 10^{-9}$ \\
    \hline
    \end{tabular}
\end{table*}

It is noteworthy that the dimensionless permeability $\K/L^2$ for the five models is not only positively
correlated with inverse of the dimensionless specific surface $(sL)^{-1}$, as shown in Ref. \cite{Ju05},
but also, as shown here, with the hyperuniformity order metric $\overline{B}_V$ (see Table
\ref{tab:Bv}), the dimensionless pore-size second moment 
$\langle{\delta^2} \rangle/L^2$ (see Table \ref{tab:moments}), and the
long-time asymptotic coefficient $C_\mathcal{S}$ of excess
spreadability $\mathcal{S}(\infty) - \mathcal{S}(t)$ [see Eq.
\eqref{eq:S_exc3}]; see the specific values summarized in Table \ref{tab:summary}.
It can be seen that the relative rankings of the five models according to all of these characteristics are the same, namely, in descending order the Schwarz P structures possess the largest values, followed by the circular-channel model, the square-channel model, the Schoen G, and the Schwarz D, the latter having the smallest values. 
Structures with a smaller dimensionless specific surface [i.e., larger
$(sL)^{-1}$] tend to possess domains of solid and void phases that are
distributed less uniformly throughout the fundamental  cell, leading to larger
values $\overline{B}_V$.
The positive correlation between $\langle \delta^2 \rangle/L^2$ and
$\K/L^2$ is consistent with the fact that a larger effective pore-channel area in a principal direction, measured
by $\pi \langle \delta^2 \rangle/L^2$, facilitates fluid transport, given the same formation factor. Similarly, the positive correlation between the coefficient $C_\mathcal{S}$ and
$\K/L^2$ is consistent with the fact that structures with larger pore regions, as measured by the dimensionless second moment $\langle \delta^2 \rangle/L^2$, takes longer for solute species to diffuse and fill the entire pore space. 
Thus, this latter correlation establishes yet another {\it cross-property} relation between the fluid permeability and diffusion properties for these 
triply periodic media but, in this case, a {\it time-dependent} transport property.
It is noteworthy that it was previously established that the {\it steady-state} mean survival time $\tau$ rank orders
the five triply periodic media models in exactly the same way as $\K$ \cite{Ge09a}.

An outstanding problem for future research is the determination of the triply periodic bicontinuous 
media within a cubic fundamental cell of side length $L$  with minimal dimensionless specific surface $sL$ for arbitrary porosities subject to cubic symmetry. 
This would be the first step in validating {\bf {\bf Conjecture \ref{conj2}}}.
Its full validation would require one to show that such optimal structures with minimal $sL$ also maximize the isotropic fluid permeability under the aforementioned constraints.

\section*{Acknowledgments}
We thank M. Skolnick for his assistance in generating some of the schematics
in Fig. \ref{5-vol-frac}.
The authors gratefully acknowledge the support of the National
Science Foundation under Award No. CBET-2133179 and the U.S. Army Research Office under Cooperative Agreement No. W911NF-22-2-0103.


\appendix
\section{Phase-Inversion Symmetry of Triply Periodic Minimal Surfaces} \label{app:phase-inversion}

Here, we prove that three triply periodic minimal surfaces (Schwarz P, Schwarz D, and Schoen G) with porosity $\phi=1/2$ possess phase-inversion symmetry \cite{To02a}, i.e., the $n$-point correlation functions, defined in Sec. \ref{n-point}, obey the following symmetry conditions:
\begin{align}	\label{eq:def_phase-inversion}
	S_n^{(1)}({\bf x}_1, ..., {\bf x}_n) = S_n^{(2)}({\bf x}_1, ..., {\bf x}_n) \quad (n=2,3,\ldots).
\end{align}
It is straightforward to prove Eq. \eqref{eq:def_phase-inversion} for the Schwarz P and D structures  by using the definition of $S_n^{(i)}$, since, for each model, the solid-phase and void-phase domains are identical under simple translations:
\begin{align}
	\text{Schwarz P}: \quad &
	{\cal I}^{(1)}({\bf r}) = {\cal I}^{(2)}({\bf r} + L/2 (\hat{\bf x} + \hat{\bf y} + \hat{\bf z})),	\label{eq:SchwarzP} \\ 
	\text{Schwarz D}: \quad &
	{\cal I}^{(1)}({\bf r}) = {\cal I}^{(2)}({\bf r} + L/2 \hat{\bf x}),	\label{eq:SchwarzD}
\end{align} 
where $\hat{\bf x}$, $\hat{\bf y}$, and $\hat{\bf z}$ are unit vectors in the directions of $x$, $y$, and $z$ axes, respectively.
For Schoen G, one can prove Eq. \eqref{eq:def_phase-inversion} by using the fact that its solid- and void-phase domains are identical under a point reflection, i.e.,
\begin{align}
	\text{Schoen G}: \quad &
	{\cal I}^{(1)}({\bf r}) = {\cal I}^{(2)}(-{\bf r} + L/2 \hat{\bf y})	\label{eq:SchoenG}.
\end{align}
Applying Eq. \eqref{eq:SchoenG} to the definition of $S_n^{(i)}$, we now prove Eq. \eqref{eq:def_phase-inversion} for Schoen G structure:
\begin{align}
	&S_n^{(1)}({\bf x}_1,\ldots, {\bf x}_n)
	\equiv
	\left \langle
		\prod_{i=1}^n 
		{\cal I}^{(1)}({\bf x}_i)
	\right \rangle
	= 
	\left \langle
		\prod_{i=1}^n 
		{\cal I}^{(2)}(-{\bf x}_i + L/2 \hat{\bf y})
	\right \rangle
	\nonumber \\
	=&
	S_n^{(2)}(-{\bf x}_1 + L/2 \hat{\bf y},-{\bf x}_2 + L/2 \hat{\bf y},\ldots, -{\bf x}_n + L/2 \hat{\bf y})
	\nonumber \\
	=&
	S_n^{(2)}(-{\bf x}_1 ,-{\bf x}_2 ,\ldots, -{\bf x}_n)
	=
	S_n^{(2)}({\bf x}_1 , {\bf x}_2 ,\ldots, {\bf x}_n) \quad \mathrm{for~} n\geq 2
	\label{eq:phase-inversion_SchoenG},
\end{align}
the last line of which implies statistically homogeneity.

\end{document}